\documentclass[prd,tightenlines,nofootinbib,superscriptaddress]{revtex4}
\usepackage{amsfonts,amssymb,amsthm,bbm}

\usepackage{amsmath,amssymb,bbm}
\usepackage{amsfonts}
\newcommand{\N}{{\mathbb N}}
\newcommand{\R}{{\mathbb R}}
\newcommand{\Z}{{\mathbb Z}}

\newcommand{\cG}{{\mathcal G}}

\newcommand{\cS}{{\mathcal S}}
\newcommand{\SU}{\mathrm{SU}}
\newcommand{\SO}{\mathrm{SO}}

\newcommand{\ISO}{\mathrm{ISO}}
\newcommand{\DSU}{\mathrm{DSU}}

\newcommand{\D}{\textrm{D}}

\newcommand{\be}{\begin{equation}}
\newcommand{\ee}{\end{equation}}
\newcommand{\beq}{\begin{eqnarray}}
\newcommand{\eeq}{\end{eqnarray}}
\newcommand{\bea}{\begin{eqnarray}}
\newcommand{\eea}{\end{eqnarray}}
\newcommand{\nn}{\nonumber}

\newcommand{\su}{{\mathfrak su}}

\newcommand{\tr}{{\mathrm Tr}}
\newcommand{\f}{\frac}
\newcommand{\tl}{\widetilde}
\newcommand{\vphi}{\varphi}

\newcommand{\Ref}[1]{(\ref{#1})}

\def\nn{\nonumber}

\def\arr{\rightarrow}

\def\ka{\kappa}
\def\tpsi{{\widetilde{\psi}}}
\def\tf{\widetilde{f}}
\def\tg{\widetilde{g}}
\def\th{\widetilde{h}}
\def\vphi{\varphi}
\def\hvphi{\widehat{\varphi}}

\begin{document}

\title{A Deformed Poincar\'e Invariance for Group Field Theories}

\author{Florian Girelli}
\affiliation{School of Physics, The University of Sydney, Sydney, New South Wales 2006, Australia}
\author{Etera R. Livine}
\affiliation{Laboratoire de Physique, ENS Lyon, CNRS-UMR 5672, 46 All\'ee d'Italie, Lyon 69007, France}

\date{\today}

\begin{abstract}
In the context of quantum gravity, group field theories are field theories that generate spinfoam amplitudes as Feynman diagrams. They can be understood as generalizations of the matrix models used for 2d quantum gravity. In particular Boulatov's theory reproduces the amplitudes of the Ponzano-Regge spinfoam model for 3d quantum gravity. Motivated by recent works on field theories on non-commutative flat spaces, we show that Boulatov's theory (and its colored version) is actually invariant under a global deformed Poincar\'e symmetry. This allows to define a notion of flat or excited  geometry states when considering  scalar perturbations around classical solutions of the group field equations of motion.
As a side-result, our analysis seems to point out that the notion of braiding of group field theories should be a key feature to study further in this context.
\end{abstract}

\maketitle




Spinfoams propose a non-perturbative framework to define a regularized path integral for quantum gravity. They allow to compute correlations for geometric observables in a 4d path integral and in particularly define transition amplitudes between quantum states of 3d geometries. They can be interpreted as the covariant space-time picture for canonical Loop Quantum Gravity and are also understood as a quantized version of (area) Regge calculus. Spinfoam models were initially constructed as state-sum models describing the evolution of (Loop Quantum Gravity's) spin network states and they are dually understood as attributing a quantum amplitude for every 4d triangulated (pseudo-)manifold interpolating between (3d) boundary states \cite{spinfoam} (see also \cite{carlo1,carlo0, fotini} for the construction of the spinfoam formulation from the loop quantum gravity formalism).

Spinfoams can then be reformulated as generalized matrix/tensor models, which have been dubbed group field theories (GFT) \cite{gft}. The triangulated space-time manifold arises as a Feynman diagram of the GFT and the evaluation of this Feynman diagram defines the corresponding spinfoam amplitude. These GFTs allow to define the non-perturbative sum over all space-time triangulations i.e over all geometries and topologies. The first GFT was introduced by Boulatov for the Ponzano-Regge spinfoam model of 3d quantum gravity \cite{boulatov}. This was shortly followed by a generalization by Ooguri to topological BF theory in four dimensions \cite{ooguri}. This approach was later applied to the Barrett-Crane model for 4d quantum gravity \cite{tous}. It was finally shown for all spinfoam models (with a local amplitude) can be derived from a GFT \cite{origin}. Since then, GFT can be considered as the fundamental formulation of spinfoam models.

Although the path integral for group field theories at the quantum level is supposed to define spinfoam models, little study has done on the classical structure of these GFTs. There have been recent works focusing on showing that the GFT path integral is well-defined at a non-perturbative level \cite{laurent,summing_laurent,summing_orsay, razvan, coloured}. However, if GFTs are to be considered as standard field theories to be quantized, we should first study them classically, define their symmetries and investigate their classical solutions in order to gain insight in their classical behavior. The goal of the present paper is to go one step forward in this direction in the context of Boulatov's group field theory for 3d quantum gravity.

In order to discuss the symmetries of the 3d group field theory, we start with the remark that spinfoam amplitudes for 3d gravity coupled to (off-shell) massive particles were shown to be equal to Feynman diagrams of a braided non-commutative quantum field theory (NCQFT) on a flat 3d background \cite{pr3,prl}. This NCQFT has been shown to be invariant under a quantum deformed Poincar\'e group \cite{pr3,prl,majid, karim, matrix}. It is only natural to investigate if this deformed Poincar\'e group also provide a symmetry of the initial GFT. Furthermore these effective NCQFTs describing the dynamics of matter field coupled to 3d quantum geometry can be defined themselves as field theories on group manifolds \cite{pr3,matrix,winston}. Actually this is a powerful viewpoint and these effective NCQFTs have been shown to arise at the classical level as specific perturbations of the original group field around particular classical solutions of the 3d GFT \cite{winston}. These perturbations seem to describe 2d geometry excitations around some non-trivial 3d background, and are called 2d phases of the 3d GFT. If the dynamics of these perturbations are invariant under Poincar\'e transformations, it is thus natural to investigate if the full theory is also invariant, and then use this as a criteria to distinguish classical solutions which conserve or break this symmetry. This viewpoint of seeing matter fields as particular 2d perturbations of the group field has also been applied to 4d spinfoam models (for BF theory) and it has been shown that it leads to non-commutative (scalar) field theories invariant under the $\ka$-deformed Poincar\'e group \cite{matter4d}. This hints towards the possibility that the Poincar\'e group also provides a relevant symmetry for 4d GFTs, even though we have decided to focus on the case of 3d GFTs in the present work.

In the first section, we quickly review the definition of the 3d GFT and its 2d phases. Then we show in the second section that the 3d GFT is actually invariant under the quantum-deformed Poincar\'e group.
We investigate the corresponding Fourier transform of the GFT in the third section.
%
Finally, in a fourth section, we use this symmetry to define ``flat" 3d geometry states which preserve this Poincar\'e  symmetry and distinguish them from ``excited" (or doped) quantum states which suppress or enhance certain gravity modes and breaks the Poincar\'e invariance. This might be used to define a notion of ``vacuum state" in this background independent framework with no clear definition of ``energy". In a last section, we start discussing the generalization of this framework to 4d GFTs.

\section{An Overview of 3d GFT and its 2d Phases}

We introduce a real field $\phi(g_1,g_2,g_3)$ on the group manifold $\SU(2)^{\times 3}$.
We assume that it satisfies a $\SU(2)$-invariance:
\be
\forall g\in\SU(2),\quad
\phi(g_1g,g_2g,g_3g)=\phi(g_1,g_2,g_3).
\ee
The action of Boulatov's group field theory is made of a trivial kinematical term and an interaction vertex representing a tetrahedron:
\be
\label{real}
S_0[\phi]=
\f12\int [dg]^3\, \phi(g_1,g_2,g_3)\phi(g_3,g_2,g_1)
-\f\lambda {4!}\int[dg]^6\, \phi(g_1,g_2,g_3)\phi(g_3,g_4,g_5)\phi(g_5,g_2,g_6)\phi(g_6,g_4,g_1).
\ee
The constant $\lambda$ is the GFT coupling constant. Each Feynman diagram of this GFT is interpreted as a 3d (pseudo-)triangulation: the field represent a triangle, the interaction vertices are tetrahedra which are glued along triangles using the trivial propagator. The evaluation of the Feynman diagram gives the spinfoam amplitude of the Ponzano-Regge model, which provides a proper quantization of 3d gravity. The coupling $\lambda$ controls the number of tetrahedra in the triangulation. After a proper rescaling of the field, it can also be seen to control the topology of the triangulation \cite{gft} (similarly to what happens with matrix models).

One could add further interaction terms, either still quartic like the pillow term \cite{laurent}, or higher order terms representing more polyhedra with more faces. In an effective field theory approach to GFT, one would eventually have to include all these terms in the action and study the flow of all the corresponding coupling constants.

One also usually considers the properties of $\phi$ under permutations of the three variables $g_1,g_2,g_3$. One can choose that $\phi$ is completely invariant under all permutations of its three arguments, or that $\phi$ is invariant under only even permutations, or that $\phi$ is not assumed to be invariant at all. If $\phi$ is not assumed invariant under permutations, then one should consider adding to the action other quartic interaction terms given by permuting some of the arguments of the fields in the tetrahedron term. In all cases, it does not change the fact that the amplitude of the Feynman diagrams give the Ponzano-Regge spinfoam amplitude. The various choices will simply lead to different statistical weights given to all triangulations \cite{tous,summing_laurent}. In the following, we will consider the form of the action given above as the fundamental one, but we will discuss the interplay between the Poincar\'e symmetry and the permutations in the next section.

Before introducing the 2d phases, we also define a complex version of the GFT. Now taking a complex field $\phi$, still $\SU(2)$-invariant , we define the following action:
\be
\label{complex}
S_c[\phi]=
\f12\int [dg]^3\, \phi(g_1,g_2,g_3)\overline{\phi(g_1,g_2,g_3)}
-\f\lambda {4!}\int[dg]^6\, \phi(g_1,g_2,g_3)\overline{\phi(g_5,g_4,g_3)}\phi(g_5,g_2,g_6)\overline{\phi(g_1,g_4,g_6)}.
\ee
Let us point out that this action is still real, since it is easy to check that $S_c=\overline{S_c}$.
The relation with the previous action is achieved by assuming that the field $\phi$ satisfies a reality condition \cite{winston}~:
\be
\overline{\phi(g_1,g_2,g_3)}=\phi(g_3,g_2,g_1).
\ee
In the following, we will not assume this condition unless stated otherwise.

The classical field equations of the GFT action $S_0$ are:
\be
\phi(g_3,g_2,g_1)=
\f\lambda{3!}\int [dg]^3\,\phi(g_3,g_4,g_5)\phi(g_5,g_2,g_6)\phi(g_6,g_4,g_1).
\ee
A class of solutions was identified in \cite{winston}. They are parameterized by an arbitrary function $f\in L^2(\SU(2))$ satisfying the normalization constraint $\int f^2=1$~:
\be
\phi_f(g_1,g_2,g_3)=\,\sqrt{\f{3!}{\lambda}}\int dg\,\delta(g_1g)f(g_2g)\delta(g_3g).
\label{flat}
\ee
These are also classical solutions for the complex action $S_c$ as long as $f$ is kept real. Up to now, no other class of classical solutions to the 3d GFT has been identified.

A 2d perturbation of the group field is defined as $\Delta\phi(g_1,g_2,g_3)\,\equiv\,\psi(g_1g_3^{-1})$. Such perturbations are obviously $\SU(2)$-invariant. Following the framework introduced in \cite{winston}, we look at the effective action for such 2d perturbations around the classical solutions:
\be
S_{eff}[\psi]\,\equiv\,
S_0[\phi_f+\psi]-S_0[\phi_f].
\ee
A tricky point is that the constant offset $S_0[\phi_f]$ is a priori an infinite constant, but this is not relevant to our discussion so we decide to put this issue aside for the moment. As shown in \cite{winston}, the remaining effective action acquires a non-trivial propagator:
\beq
\label{eff}
S_{eff}^{(f)}[\psi]&=&
\f12\int dg\,K(g)\psi(g)\psi(g^{-1})
-\sqrt{\f{\lambda}{3!}}\left(\int f\right) \int [dg]^3\,\psi(g_1)\psi(g_2)\psi(g_3)\delta(g_1g_2g_3) \nn\\
&&-\f{\lambda}{4!}\int [dg]^4\,\psi(g_1)\psi(g_2)\psi(g_3)\psi(g_4)\delta(g_1g_2g_3g_4),
\eeq
where the quadratic term is given by:
\be
K(g)\,\equiv\,1-\left(\int f\right)^2-\int dh\,f(h)f(hg).
\ee
If the parameter $f$ is chosen such that $\int f=0$, then the extra mass term in the propagator and the cubic interaction term drop out. Such examples are provided by the characters $\chi_j(g)$ of the irreducible representations of $\SU(2)$ labeled by the spin  $j\in\N/2$, which provide a orthonormal basis of $L^2$ central functions on $\SU(2)$ and which satisfy both conditions $\int (\chi_j)^2=1$ and $\int \chi_j =0$.

A special case of this construction is the case of the trivial classical solution $f=0$, $\phi_f=0$. This amounts to directly restricting  the group field to its 2d perturbation, $\phi(g_1,g_2,g_3)\,\equiv\,\psi(g_1g_3^{-1})$ and leads to the standard 2d group field theory (which has been shown to be equivalent to the usual matrix models)~:
\be
\label{2d}
S_0[\phi=\psi]\,=\,\f12\int dg\,\psi(g)\psi(g^{-1})
-\f{\lambda}{4!}\int [dg]^4\,\psi(g_1)\psi(g_2)\psi(g_3)\psi(g_4)\delta(g_1g_2g_3g_4).
\ee
This defines the 2d regime of the 3d GFT. Let us underline an important issue. If we restrict the group field to a different choice of 2d perturbations, we lose the nice structure of the effective action. Indeed, let us instead choose $\phi(g_1,g_2,g_3)\,\equiv\,\tpsi(g_1g_2^{-1})$. Then the corresponding 2d regime is described by the action:
\be
S_0[\phi=\tpsi]\,=\,\f12\left(\int \tpsi\right)^2
-\f{\lambda}{4!}\left(\int \tpsi\right)^4.
\ee
This is a theory of a single real variable. This issue is actually cured by the complex action \Ref{complex} which we proposed above. The three different choices of 2d perturbations simply lead to a different ordering of the fields in the interaction term:
\beq
S_c[\phi=\psi(g_1g_3^{-1})]&=&\f12\int dg\,\psi(g)\overline{\psi(g)}
-\f{\lambda}{4!}\int [dg]^4\,\psi(g_1)\overline{\psi(g_2)}\psi(g_3)\overline{\psi(g_4)}\,\delta(g_1(g_2)^{-1}g_3(g_4)^{-1}), \label{2dc1}\\
S_c[\phi=\psi(g_1g_2^{-1})]&=&\f12\int dg\,\psi(g)\overline{\psi(g)}
-\f{\lambda}{4!}\int [dg]^4\,\psi(g_1)\psi(g_2)\overline{\psi(g_3)}\,\overline{\psi(g_4)}\,\delta(g_1(g_2)^{-1}g_3(g_4)^{-1}),
\label{2dc2}\\
S_c[\phi=\psi(g_2g_3^{-1})]&=&\f12\int dg\,\psi(g)\overline{\psi(g)}
-\f{\lambda}{4!}\int [dg]^4\,\psi(g_1)\overline{\psi(g_2)}\,\overline{\psi(g_3)}\psi(g_4)\,\delta(g_1(g_2)^{-1}g_3(g_4)^{-1}).
\label{2dc3}
\eeq

The last point of this section deals with the symmetries of these 2d phases. The field theory actions \Ref{eff}, \Ref{2d} and \Ref{2dc1} are all invariant under the quantum double $\DSU(2)$, which provides a quantum deformation of the Euclidean 2d Poincar\'e group $\ISO(3)$.
The cases of actions \Ref{2dc2} and \Ref{2dc3} are more subtle since the inverse operators ${}^{-1}$ seem not to be consistently matched with the complex conjugations. Nevertheless, due to the braiding of the theory, these two actions can be shown to be equivalent to the first one \Ref{2dc1}, as we will explain later in this section.
These actions are actually written in the momentum representation: the momentum space is the homogeneous space $\SU(2)\sim \cS_3$, it is curved and thus the space-time constructed as the dual coordinate space is non-commutative. The conservation of momentum is implemented in the actions by the constraints $\delta(g_1..g_n)$ both in the kinematical and interaction terms.
As explained in \cite{pr3,prl,majid, karim, matrix}, the Fourier transform is defined by the plane waves $\exp(\tr\, x g)$ with the coordinate vector $x\in\R^3\sim\su(2)$, the group momentum $g\in\SU(2)$ and the trace $\tr$ taken in the fundamental two-dimensional representation. These plane waves can be re-written as $\exp(i\vec{x}\cdot\vec{p}(g))$, where the momentum vector $\vec{p}(g)\equiv\f1{2i}\tr\,g\,\vec{\sigma}$ is the projection of the group element $g$ on the Pauli matrices and defines (stereographic) coordinates on the 3-sphere (divided by $\Z_2$). A discussion on other possible choices of momentum coordinate can be found in \cite{karim,ryan}.
Rotations are parameterized by group elements $\Lambda\in\SU(2)$ act on the field by conjugation:
\be
\psi(g)\,\arr\,\psi(\Lambda^{-1}g\Lambda),\quad
\overline{\psi(g)}\,\arr\,\overline{\psi(\Lambda^{-1}g\Lambda)}.
\ee
Translations are parameterized by $x\in\su(2)\sim\R^3$ and act by multiplication on the field by the phase $\exp(\tr\, x g)=\exp(i\vec{x}\cdot\vec{p}(g))$:
\be
\psi(g)\,\arr\,e^{\tr\, x g}\,\psi(g),\quad
\overline{\psi(g)}\,\arr\,\overline{e^{\tr\, x g}\psi(g)}=e^{-\tr\, x g}\,\overline{\psi(g)}=e^{\tr\, x g^{-1}}\,\overline{\psi(g)}.
\ee
The non-commutativity is encoded in the non-trivial co-product of the quantum deformation, that is in the action of translations on (tensor) product of the field:
\be
\psi(g_1)\psi(g_2)\,\arr\,e^{\tr\, x g_1g_2}\,\psi(g_1)\psi(g_2),\quad
\psi(g_1)\overline{\psi(g_2)}\,\arr\,e^{\tr\, x g_1(g_2)^{-1}}\,\psi(g_1)\overline{\psi(g_2)},
\quad \dots
\ee
This in turn leads to a non-commutative $\star$-product between plane waves and a non-commutative addition $\oplus$ of momentum, defined such that:
\be
e^{i\vec{x}\cdot\vec{p}_1}\star e^{i\vec{x}\cdot\vec{p}_2}
\,=\,
e^{\tr\, x g_2}\star e^{\tr\, x g_2}
\,=\,
e^{\tr\, x g_1g_2}
\,=\,
e^{i\vec{x}\cdot(\vec{p}_1\oplus\vec{p}_2)},
\ee
with $\vec{p}_k\equiv\vec{p}(g_k)$. Details on this construction can be found in \cite{pr3,majid,karim}. What interests us in the present paper is that the 3d GFT is actually invariant under this same deformed Poincar\'e symmetry, as we show in the next section.

Before moving to the 3d GFT, we need to comment on the braiding of these non-commutative field theories. Looking at the action of the translations on the field $\psi$, it is clear that the action on (tensor) products of two field insertions is not symmetric under the exchange of these two fields. More precisely, $\psi(g_1)\psi(g_2)$ and $\psi(g_2)\psi(g_1)$ do not transform the same way since the first one is multiplied by the phase $\exp(\tr\, x g_1g_2)$ and the later by the phase $\exp(\tr\, x g_2g_1)$. The proper way to exchange the two field insertions is to introduce a non-trivial braiding (see e.g. \cite{pr3,prl} for more details)~:
\be
\psi(g_1)\psi(g_2)\,\arr\,\psi(\tl{g}_2)\psi(\tl{g}_1)
\qquad
\textrm{with}
\quad
\tg_1=g_2^{-1}g_1g_2,\quad
\tg_2=g_2,\quad
\tg_2\tg_1=g_1g_2\ne g_2g_1.
\ee
Then $\psi(g_1)\psi(g_2)$ and $\psi(\tl{g}_2)\psi(\tl{g}_1)$ both transform exactly the same way under the deformed Poincar\'e translations.
We can apply this simple reasoning to the interaction term of the 2d actions. Starting with the real field, the interaction term is the integral of the density $\psi(g_1)..\psi(g_4)\,\delta(g_1g_2g_3g_4)$ where the order of the group elements $g_1$,..,$g_4$ is important in the $\delta$-function. The braiding allows to switch this order. For example, applying the braiding between the first and second field insertions give:
\be
\int \psi(g_1)\psi(g_2)..\,\delta(g_1g_2g_3g_4)
\,=\,
\int \psi(\tg_2^{-1}\tg_1\tg_2)\psi(\tg_2)..\,\delta(\tg_2\tg_1g_3g_4)
\,=\,
\int \psi(\tg_1)\psi(\tg_2)..\,\delta(\tg_2\tg_1g_3g_4),
\ee
where the second equality follows from the left and right invariance of the Haar measure on $\SU(2)$.
We can extend this logic to the case of a complex field and show that the three actions of the 2d phases, \Ref{2dc1} to \Ref{2dc3}, are equivalent. For instance, to go from \Ref{2dc2} to \Ref{2dc1}, we need to introduce the following braiding between insertions of $\psi$ and $\overline{\psi}$~:
\be
\psi(g_1)\overline{\psi(g_2)}\,\arr\,\overline{\psi(\tg_2)}\psi(\tl{g}_1),\qquad
\textrm{with}
\quad
\tg_1=g_1^{-1},
\,
\tg_2=g_1^{-1}g_2^{-1}g_1,
\quad
\tg_2^{-1}\tg_1=g_1^{-1}g_2.
\ee
Then using the invariance properties of the Haar measure, it is straightforward to conclude that the apparently different interaction terms are actually all equal, by first swapping $g_2$ and $g_3$ for \Ref{2dc2} and then swapping $g_3$ and $g_4$ for \Ref{2dc3}. This shows that the three 2d phases are all equivalent and thus all invariant under the quantum-deformed Poincar\'e group $\DSU(2)$.

\section{Poincar\'e Invariance for 3d Group Field Theory}

Following the results reviewed in the previous section, it is fairly straightforward to see that the action of the 3d GFT itself is invariant under the deformed Poincar\'e transformations. The natural variables to consider are not the original arguments of the field $g_1,g_2,g_3$ but the gauge invariant combinations $g_1g_2^{-1},g_2g_3^{-1},g_1g_3^{-1}$ as used to define the 2d perturbations of the group field $\phi$.

Indeed, the action $S_0[\phi]$  given in \Ref{real} for a real field is clearly invariant under the Poincar\'e transformations on $g_1g_3^{-1}$, with rotations still parameterized by $\Lambda\in\SU(2)$ and translations by $x\in\su(2)$~:
\be
\left|
\begin{array}{rcl}
\phi(g_1,g_2,g_3)&\arr&\phi(\Lambda g_1,g_2,\Lambda g_3), \\
\phi(g_1,g_2,g_3)&\arr&e^{\tr\, x g_1g_3^{-1}}\,\phi(g_1,g_2,g_3).
\end{array}
\right.
\ee
Let us notice that these $\SU(2)$ transformations reproduce as expected the adjoint action (by conjugation) of $\SU(2)$ on the variable $g_1g_3^{-1}$ and thus lead to the correct $\SU(2)$ transformation for the 2d perturbation $\psi(g_1g_3^{-1})$.
This   action $S_0[\phi]$  is only invariant under translations with respect to this variable $g_1g_3^{-1}$ and is not invariant under translations along the two other gauge invariant combinations. Let us nevertheless point out that the translations are not rigourously defined as above since we are acting with a complex phase on a real field. Thus it is better to work right from the start with a complex field.

The action $S_c[\phi]$ for a complex field introduced in \Ref{complex} is similarly invariant  deformed Poincar\'e transformations:
\be\label{rotation}
\left|
\begin{array}{lcl}
\phi(g_1,g_2,g_3)\,\arr\,\phi(\Lambda g_1,g_2,\Lambda g_3),
&\quad&
\overline{\phi(g_1,g_2,g_3)}\,\arr\,\overline{\phi(\Lambda g_1,g_2,\Lambda g_3)} \\
\phi(g_1,g_2,g_3)\,\arr\,e^{\tr\, x g_1g_3^{-1}}\,\phi(g_1,g_2,g_3),
&\quad&
\overline{\phi(g_1,g_2,g_3)}\,\arr\,e^{\tr\, x g_3g_1^{-1}}\,\overline{\phi(g_1,g_2,g_3)}.
\end{array}
\right.
\ee
These transformations are consistent with the reality condition $\overline{\phi(g_1,g_2,g_3)}=\phi(g_3,g_2,g_1)$ since these two fields have the same transformation laws.
The action $S_c[\phi]$ is not obviously invariant under translations along the two other gauge invariant combinations. This seems to be related to the braiding issue discussed at the end of the previous section, thus there might be some non-trivial braiding of the group field $\phi$ which would allow to show that $S_c[\phi]$ is invariant under the two other types of deformed Poincar\'e transformations. However, we haven't been able to find such a mechanism yet.

This issue is actually deeper: if we change the ordering of the arguments of the fields in these actions $S_0$ and $S_c$, then we lose the invariance under the (deformed) Poincar\'e transformations. Actually it is straightforward to see that the fact that there exists a consistent 2d regime with effective action \Ref{2d} is equivalent to the existence of a Poincar\'e symmetry. A way out of this problem is move to a multi-scalar group field theory or colored group field theory as already introduced  in \cite{coloured}. This solves both the issues of having the three types of Poincar\'e transformations for the standard ordering of the arguments usually used and of generalizing this to an arbitrary ordering. These colored group field theories have the same evaluations of Feynman diagrams up to statistical factors (symmetry factors of the Feynman diagrams). Thus, from a mathematical point of view, using a colored GFT seems natural since it reproduces the same spinfoam amplitudes as the standard GFT with a single group field and only the statistical weights associated to each space-time triangulation are modified. However, from a physical point of view, if we consider the GFT as the fundamental formulation for spinfoam models, then introducing more scalar fields would require a physical interpretation.

Indeed, first let us consider the colored generalization of the original complex action $S_c$. We now work with four fields $\phi_1,..,\phi_4$ and define:
\be
\label{colored0}
S_c^{col}[\phi]=
\f12\int [dg]^3\, \sum_i\phi_i(g_1,g_2,g_3)\overline{\phi_i(g_1,g_2,g_3)}
-\f\lambda {3!}\int[dg]^6\, \phi_1(g_1,g_2,g_3)\overline{\phi_2(g_5,g_4,g_3)}\phi_3(g_5,g_2,g_6)\overline{\phi_4(g_1,g_4,g_6)}.
\ee
Then the three types of Poincar\'e transformations are:
\be
\left|
\begin{array}{rcl}
\phi_i(g_1,g_2,g_3)&\arr&\phi_i(\Lambda g_1,g_2,\Lambda g_3), \\
\phi_1(g_1,g_2,g_3)&\arr&e^{\tr\, x g_1g_3^{-1}}\,\phi_1(g_1,g_2,g_3),\\
\phi_2(g_1,g_2,g_3)&\arr&e^{\tr\, x g_1g_3^{-1}}\,\phi_2(g_1,g_2,g_3),\\
\phi_3(g_1,g_2,g_3)&\arr&e^{\tr\, x g_1g_3^{-1}}\,\phi_3(g_1,g_2,g_3),\\
\phi_4(g_1,g_2,g_3)&\arr&e^{\tr\, x g_1g_3^{-1}}\,\phi_4(g_1,g_2,g_3)
\end{array}
\right.
\ee
\be
\left|
\begin{array}{rcl}
\phi_i(g_1,g_2,g_3)&\arr&\phi_i(\Lambda g_1,\Lambda g_2,g_3), \\
\phi_1(g_1,g_2,g_3)&\arr&e^{\tr\, x g_1g_2^{-1}}\,\phi_1(g_1,g_2,g_3),\\
\phi_2(g_1,g_2,g_3)&\arr&e^{\tr\, x g_2g_1^{-1}}\,\phi_2(g_1,g_2,g_3),\\
\phi_3(g_1,g_2,g_3)&\arr&e^{\tr\, x g_2g_1^{-1}}\,\phi_3(g_1,g_2,g_3),\\
\phi_4(g_1,g_2,g_3)&\arr&e^{\tr\, x g_1g_2^{-1}}\,\phi_4(g_1,g_2,g_3)
\end{array}
\right.
\ee
\be
\left|
\begin{array}{rcl}
\phi_i(g_1,g_2,g_3)&\arr&\phi_i(g_1,\Lambda g_2,\Lambda g_3), \\
\phi_1(g_1,g_2,g_3)&\arr&e^{\tr\, x g_2g_3^{-1}}\,\phi_1(g_1,g_2,g_3),\\
\phi_2(g_1,g_2,g_3)&\arr&e^{\tr\, x g_2g_3^{-1}}\,\phi_2(g_1,g_2,g_3),\\
\phi_3(g_1,g_2,g_3)&\arr&e^{\tr\, x g_3g_2^{-1}}\,\phi_3(g_1,g_2,g_3),\\
\phi_4(g_1,g_2,g_3)&\arr&e^{\tr\, x g_3g_2^{-1}}\,\phi_4(g_1,g_2,g_3)
\end{array}
\right.
\ee
The first case is the Poincar\'e symmetry already written for the case of a single field. As for the other cases, the quadratic kinetic term is trivially invariant. However, the interaction term require a specific ordering of the field, $\phi_1\phi_3\overline{\phi_2}\,\overline{\phi_4}$ for the translations along $g_1g_2^{-1}$ and $\phi_1\overline{\phi_2}\,\overline{\phi_4}\phi_3$ for the translations along $g_2g_3^{-1}$. Thus, we either have to swap the field insertions $\phi_3$ with $\overline{\phi_2}$ or $\overline{\phi_4}$ with $\phi_3$. In both cases, these are the fields which have their behavior under translations inverted compared to the standard field transformations. This is a hint towards a possible non-trivial braiding of the original group field $\phi$. We postpone the analysis of this issue to future work, focusing instead in the present paper on the role of the deformed Poincar\'e symmetry for the GFT.

Finally, as we said earlier, using this colored group field theory, we can now allow for any ordering of the arguments $g_1,..,g_6$ of the fields in the interaction term and postulate accordingly new transformation rules of the fields $\phi_1,..,\phi_4$, and we will always identify the three types of Poincar\'e symmetries.

\medskip

Since we have field transformations leading to a symmetry of the action, the natural question is: what are the Noether current and charges and the corresponding Ward identities at the quantum level? First, since there is no derivative in the action, especially no time derivative, and that the kinematic term is trivial (trivial propagator), the Noether construction only gives a trivial answer. Second, unsurprisingly, the Ward identities simply state that the expectation values of observables are invariant under these Poincar\'e transformations.

\section{Gauge Reduction and Fourier Transform}

Discussing a momentum representation and a Poincar\'e symmetry for the group field theory, it is natural to discuss its Fourier transform and a space-time representation for the GFT. Indeed, although the GFT a priori provides us with a background-independent non-perturbative formulation for spinfoam models, we can see the GFT as a field theory defined on a flat non-commutative 3d space-time. As we explain below, this comes as a generalization of the Fourier transform mapping the 2d GFTs to 3d non-commutative field theories \cite{matrix}.

In order to see this clearly, we start by a gauge reduction of the group field. Indeed, although the group field $\phi(g_1,g_2,g_3)$ is a priori a function of three group elements, it effectively depends on only two arguments due to its gauge invariance:
\be
\phi(g_1,g_2,g_3)=\phi(g_1g_3^{-1},g_2g_3^{-1},1)\,\equiv\,\vphi(g_1g_3^{-1},g_2g_3^{-1}).
\ee
We can reformulate the 3d GFT in term of this gauge-fixed field $\vphi$:
\be
S_c[\phi]=S[\vphi]\equiv\,
\f12\int [dG]^2\, \vphi(G_1,G_2)\overline{\vphi(G_1,G_2)}
-\f\lambda {4!}\int[dG]^5\, \vphi(G_1,G_2)\overline{\vphi(G_5,G_4)}\vphi(G_5G,G_2G)\overline{\vphi(G_1G,G_4G)},
\ee
where we did the changes of variable $G_i=g_ig_3^{-1}$ for $i=1,2,4,5$ and $G=g_3g_6^{-1}$. Written as such, the Feynman diagrams of the GFT lose their obvious geometrical interpretation in terms of triangles and tetrahedra. Nevertheless, it makes it clearer to see the Poincar\'e invariance. Indeed, the action $S[\vphi]$ is invariant under deformed translations acting on the first argument of the field $\vphi$~:
\be
\vphi(G_1,G_2)\,\arr\, e^{\tr x G_1}\,\vphi(G_1,G_2),
\ee
while it is easy to see that the action will not be invariant under translations acting on the second argument defined the same way. As we said above, this issue might be solved by some appropriate braiding, but we postpone this question to future study. A further subtlety arises when writing down the rotations acting on the field $\vphi$. The action is naturally invariant under the left-action of $\SU(2)$ on the first argument, $\vphi(G_1,G_2)\,\arr\, \vphi(\Lambda G_1,G_2)$. However, this does not correspond to the standard action of the rotations, which should act by conjugation. The action $S[\vphi]$ is not invariant under conjugation on the first argument of $\vphi$ but it requires acting on the second argument as well:
\be
\vphi(G_1,G_2)\,\arr\, \vphi(\Lambda G_1\Lambda^{-1},G_2\Lambda^{-1}).
\ee
This transformation is the $\SU(2)$ transformation induced from \eqref{rotation}.
Since we have a Poincar\'e invariance, it is natural to look at the Fourier transform of $\vphi$ on the first argument. Following the previous works \cite{pr3,prl,majid,karim,matrix}, we introduce the Fourier field:
\be
\hvphi(x,G)=\int dG_1\,e^{\tr x G_1}\,\vphi(G_1,G).
\ee
Under the assumption of working with an even field, satisfying $\vphi(G_1,G_2)=\vphi(-G_1,G_2)$ (see \cite{majid,karim,matrix} for more details), we can re-write the action $S[\vphi]$ in term of this Fourier transformed field:
$$
S[\vphi]=
\f12\int dxdG\, \hvphi(x,G)\overline{\hvphi(x,G)}
-\f\lambda {4!}\int dxdy[dG]^3\,
\left(\hvphi(x,G_2)\star e^{\tr x G}\star\overline{\hvphi(x,G_4G)}\right)
\left(e^{\tr y G^{-1}}\star\overline{\hvphi(y,G_4)}\star \hvphi(y,G_2G)\right).
$$
This gives a representation of Boulatov's GFT for 3d quantum gravity as a Poincar\'e invariant field theory over the non-commutative 3d space. This action is definitely strange, and we haven't been able to find some natural interpretation. Moreover, the interpretation of the second argument of the field remains obscure: does it represent some internal degree of freedom or could it seen as some extra-dimensions? Since a field with a single argument defines a 2d GFT which generates 2d triangulations as Feynman diagrams, the second argument of the field here should generate the third dimension. But this can not be seen easily in the formulation above.
At the end of the day, maybe we should use a different Fourier transform which might involve $G_2$ as well as $G_1$ and which might lead to a simpler field theory action.


%
%
%
%


\section{Classical GFT Solutions and Flat/Excited States of 3d Geometry}

It is interesting to look at the classical solutions of the 3d GFT from the point of view of this new Poincar\'e symmetry. Indeed, as we have already seen, the (effective) field theories describing 2d perturbations around a classical solution $\phi_f$ (defined in \Ref{flat} in the first section) are all invariant under the deformed Poincar\'e transformations. This can be easily shown from the fact that this class of classical backgrounds $\phi_f$ are invariant under Poincar\'e translations~\footnotemark:
\be
\phi_f(g_1,g_2,g_3)=\,\sqrt{\f{3!}{\lambda}}\int dg\,\delta(g_1g)f(g_2g)\delta(g_3g)
\,\arr\quad
e^{\tr\, xg_1g_3^{-1}}\phi_f(g_1,g_2,g_3)=\phi_f(g_1,g_2,g_3),
\ee
since these classical solutions impose that the group elements $g_1$ and $g_3$ are equal.
\footnotetext{
It is also trivial to check that this ansatz is also invariant under Lorentz rotations and thus under all Poincar\'e transformations:
$$
\phi_f(g_1,g_2,g_3)=\,\sqrt{\f{3!}{\lambda}}\int dg\,\delta(g_1g)f(g_2g)\delta(g_3g)
\,\arr\quad
\phi_f(\Lambda g_1,g_2,\Lambda g_3)=\phi_f(g_1,g_2,g_3).
$$
}
Since these solutions are translational-invariant, the physics of the perturbations around them is naturally Poincar\'e-invariant. Thus we call them {\it flat solutions} to the 3d GFT.

Now, we will introduce a new family of classical solutions, which break the Poincar\'e invariance, and we will call them {\it doped solutions}. The new ansatz is given the group-averaged product of three $\SU(2)$ characters probably normalized in order to satisfy the classical field equations:
\be
\phi_{(j_1,j_2,j_3)}(g_1,g_2,g_3)=\,\f{\sqrt{d_{j_1}d_{j_2}d_{j_3}}}{|\{6j\}|}\,\sqrt{\f{3!}{\lambda}}\int dg\,\chi_{j_1}(g_1g)\chi_{j_2}(g_2g)\chi_{j_3}(g_3g).
\ee
The character $\chi_j(g)$ is simply the trace of the group element $g$ in the irreducible representation of spin $j$. They satisfy the following simple convolution property useful for integral calculations:
$$
\int dg \,\chi_j(gh)\chi_k(g\th)
\,=\,
\f{\delta_{jk}}{d_j}\chi_j(h^{-1}\th).
$$
The factors $d_j=(2j+1)$ is the dimension of the irrep of spin $j$. Finally, the 6j-symbol is the standard invariant from $\SU(2)$ representation theory. Here it is more convenient to give the integral formula for the square of the 6j-symbol:
\be
\{6j\}^2=
\left\{
\begin{array}{ccc}
j_1 & j_2 & j_3 \\
j_1 & j_2 & j_3
\end{array}
\right\}^2
\,=\,
\int [dhdadbdc]\,
\chi_{j_1}(ha^{-1})\chi_{j_2}(hb^{-1})\chi_{j_3}(hc^{-1})
\chi_{j_1}(bc^{-1})\chi_{j_2}(ca^{-1})\chi_{j_3}(ab^{-1}).
\ee
We assume of course that the three irreps $j_1,j_2,j_3$ are compatible, i.e that they satisfy the triangular inequalities and that $(j_1+j_2+j_3)$ is an integer.
To check that this ansatz correctly provides a classical solution satisfying the field equation,
$$
\overline{\phi(g_1,g_2,g_3)}=
\f\lambda{3!}\int [dg]^3\,\overline{\phi(g_5,g_4,g_3)}\phi(g_5,g_2,g_6)\overline{\phi(g_1,g_4,g_6)},
$$
we only need the following identity:
$$
\forall g_i,\,
\int [dadbdc]\,\chi_{j_1}(g_1a^{-1})\chi_{j_2}(g_2b^{-1})\chi_{j_3}(g_3c^{-1})
\chi_{j_1}(bc^{-1})\chi_{j_2}(ca^{-1})\chi_{j_3}(ab^{-1})
=\{6j\}^2\,\int dg\,\chi_{j_1}(g_1g)\chi_{j_2}(g_2g)\chi_{j_3}(g_3g).
$$
This identity holds because there exists a unique left and right invariant function in $L^2(\SU(2)^{\times 3})$, which corresponds to the unique 3-valent intertwiner between the irreps $j_1,j_2,j_3$, which is given simply by the associated Clebsh-Gordan coefficients. The normalization factor $\{6j\}^2$ is computed by calculating the scalar product between the two functions on the left and right hand sides of the equation.

This new solution $\phi_{(j_1,j_2,j_3)}(g_1,g_2,g_3)$ has a simple Fourier transform under decomposition onto $\SU(2)$ representations: it unsurprisingly corresponds to the single Clebsh-Gordan coefficient between the irreps $j_1,j_2,j_3$. And as, we well know, the Clebsh-Gordan coefficients are invariant under 3-1 Pachner move up to a $\{6j\}$ factor.

\medskip

This new type of classical solutions is obviously not invariant under translations since the group elements $g_1$ and $g_3$ are not identified anymore. Thus we do not expect the effective field theory describing the dynamics of 2d perturbations around them to be Poincar\'e invariant. We define as before the effective action for the field $\psi$:
\be
S^{(j_i)}_{eff}[\psi]\,\equiv\,
S_c[\phi_{(j_1,j_2,j_3)}+\psi(g_1g_3^{-1})]-S_c[\phi_{(j_1,j_2,j_3)}].
\ee
A first remark is that contrary to the classical solutions $\phi_f$, the constant off-shift $S_c[\phi_{(j_1,j_2,j_3)}]$ is now finite:
\be
S_c[\phi_{(j_1,j_2,j_3)}]\,=\,\f32\,\f{1}{\lambda\{6j\}^2}.
\ee
Now, we can straightforwardly compute the effective action, keeping only the quadratic, cubic and quartic terms since the constant has been subtracted and the linear term vanishes. Here we focus on the quadratic terms, which define the propagator of the theory:
\beq
S^{(j_i)}_{eff}[\psi]&=&
\f12\int dg\,\psi(g)\overline{\psi(g)}
-\delta_{{j_2},0}\delta_{{j_1},{j_3}}d_{j_1}^2\int dgdh\,\psi(g)\overline{\psi(hg)}\chi_{j_1}(h)\\
&&-\f{d_{j_1}d_{j_3}}{4\{6j\}^2}\int [dg]^5
\,\left[\psi(g_1)\psi(g_2)\chi_{j_1}(g_1a)\chi_{j_1}(g_2b)\chi_{j_2}(ba^{-1}hg_2)\chi_{j_3}(hg_2a^{-1})\chi_{j_3}(hg_2b))+c.c.\right]
+\dots\nn
\eeq
This new effective action clearly looks more complicated that the previous one $S^{(f)}_{eff}$ describing 2d perturbations around flat solutions. It is easy to see that the new action, which still describes the dynamics of a scalar field on the non-commutative $\R^3$ space, is not Poincar\'e invariant and violates the conservations of momentum. For instance, the second term of the propagator allows a momentum $g$ to propagate into a momentum $hg$ with a probability amplitude given by the character $\chi_{j_1}(h)$. Heuristically, it seems that the spin labels $j_1,j_2,j_3$ defines the length scale at which the Poincar\'e symmetry is violated. In order to prove such a statement, we would need to compute explicitly the Fourier transform of the effective field action.
Finally, let us nevertheless point out that this effective field theory is still Lorentz-invariant.

\medskip

We interpret this momentum violation as the scalar matter field $\psi$ interacting strongly with the background gravitational field, with gravity pumping energy/momentum into the field thus affecting the momenta of particles. From this point of view, it seems natural to interpret these doped classical solutions as excited states of the gravitational field, whereas flat solutions seem to describe vacuum states of the gravitational field on which the dynamics of matter fields is still Poincar\'e invariant. Therefore, we would like to propose to use this new Poincar\'e symmetry for group field theory to define a notion of vacuum states of the group field theory and distinguish them from excited states where the geometry would interact strongly with the propagating matter.

This seems to work for the GFT for 3d gravity. In the four-dimensional case, things will be more complicated, since we do not expect the 4d GFT for gravity to be Poincar\'e invariant and the interaction between matter and geometry is more complex.

\section{Generalization to 4d GFTs}

We can generalize our framework to the four-dimensional case. We do not discuss group field theories for 4d gravity, but start by studying the simpler group field theories for topological BF theory (Ooguri model). Considering an arbitrary Lie group $\cG$, we introduce a  field $\vphi$ living on $\cG^{\times 4}$ satisfying the following invariance under the diagonal right action of the group:
\be
\forall g_i,g\in\cG,\,\vphi(g_1,g_2,g_3,g_4)=\vphi(g_1g,g_2g,g_3g,g_4g).
\ee
The 4d GFT action then reads as:
\beq\label{ooguri}
S[\vphi]&=&
\f12\int [dg]^4\,\vphi(g_1,g_2,g_3,g_4)\vphi(g_4,g_3,g_2,g_1) \\
&&-\f{\lambda}{5!}\int[dg]^{10}\,\vphi(g_1,g_2,g_3,g_4)\vphi(g_4,g_5,g_6,g_7)\vphi(g_7,g_3,g_8,g_9)\vphi(g_9,g_6,g_2,g_{10})\vphi(g_{10},g_8,g_5,g_1). \nn
\eeq
The reality condition on the field is:
$$
\vphi(g_4,g_3,g_2,g_1)=\overline{\vphi(g_1,g_2,g_3,g_4)},
$$
and we can introduce further complex actions by replacing $\vphi$  in the action above by their complex conjugate $\overline{\vphi}$ as long as we reverse the order of the arguments of the field, $\vphi(g_a,g_b,g_c,g_d)\arr\overline{\vphi(g_d,g_c,g_b,g_a)}$.

It is easy to see that this action is invariant under the double $\D\cG$ of the group $\cG$  acting  as in the 3d case:
\be
\vphi(g_1,g_2,g_3,g_4)\arr \vphi(\Lambda g_1,g_2,g_3,Gg_4),\quad \Lambda\in \cG, \qquad
\vphi(g_1,g_2,g_3,g_4)\arr e^{\tr \,Xg_1g_4^{-1}}\vphi(g_1,g_2,g_3,g_4),
\ee
with the same definition of the $\star$-product between plane waves:
\be
e^{\tr \,Xg_1}\star e^{\tr \,Xg_2}\equiv e^{\tr \,Xg_1g_2}.
\ee
The first type of action is interpreted as rotations, while the second set of transformations is understood as translations. Nevertheless, the double $\D\cG$ is in general not a deformation of a Poincar\'e group. In the special case where $\cG=\SU(2)$ as in the 3d case, then $\DSU(2)$ is the quantum deformation of the 3d Poincar\'e group $\ISO(3)$, but this is a special coincidence. To determine the most general (quantum) group of symmetry of the group field theory \eqref{ooguri}, given the group $\cG$ is actually an interesting question which we leave for further investigations.

We can also identify a family of flat classical solutions, which are translational-invariant:
\be
\vphi_{f,\tf}(g_1,g_2,g_3,g_4)\equiv
{}^3\sqrt{\f{4!}{\lambda}}\,\int dg\,\delta(g_1g)f(g_2g)\tf(g_3g)\delta(g_4g),
\ee
with the normalization constraint $(\int f\tf)^3=1$. The 2d perturbations will then be invariant under $\D\cG$.
Alternatively, there also exists another set of $\D\cG$ transformations acting on the sector $g_2g_3^{-1}$ which leave the GFT action invariant and we can similarly introduce flat classical solutions with respect to these transformations and the corresponding 2d perturbations.


Finally, in order to obtain (4d) GFTs which are invariant under (a deformation of) the 4d Poincar\'e group, we could start with a GFT invariant under $\D\cG$ transformations as above and break the symmetry down to the Poincar\'e group (provided $\cG$ is large enough). Following the ideas of \cite{matter4d} where some phases of the 4d GFT for $\SO(4,1)$ was shown to be invariant under the $\ka$-deformation of the Poincar\'e group $\ISO(3,1)$, such a symmetry breaking could be done either by hand by considering a GFT on a coset space (like $\SO(4,1)/\SO(3,1)$ used in \cite{matter4d}) or by modifying the GFT action to reduce its symmetries.
We postpone a detailed study of these possibilities to future investigation.

\section{Outlook}

To conclude, we have identified a (quantum deformed) Poincar\'e invariance for 3d GFTs. This provides a natural space-time interpretation for GFTs, if we interpret the group field $\phi(g_1,g_2,g_3)$ as the momentum representation and the group manifold as momentum space. This interpretation is supported by the case of 2d GFTs which are naturally mapped to non-commutative QFTs (see e.g. \cite{matrix}).

An implication of this point of view is that introducing a non-trivial propagator for the GFT does not necessarily requires inserting derivative operators acting of the group field $\phi(g_1,g_2,g_3)$ but more simply insert a non-trivial gauge-invariant function $\ka(g_1,g_2,g_3)$ in the kinetic term $\int \ka\phi\overline{\phi}$ with $\ka(g_1,g_2,g_3)$ playing the role of the usual $p^2+m^2$ of standard QFT.

We have used this new Poincar\'e invariance to discriminate between the standard flat classical solutions to the GFT which respect this Poincar\'e invariance and doped classical solutions which break this Poincar\'e invariance. Looking at perturbations around these flat solutions lead to effective matter field which are invariant under Poincar\'e transformations, while perturbations around doped solutions lead to field theories with anomalies.

Now this deformed Poincar\'e invariance of the GFT opens the door to many questions. Since GFTs can be understood as a class of non-commutative field theories, we should investigate the issues of statistics and braiding of the group field. We should also study the relation between the dual flat space-time associated to these Poincar\'e transformations and the actual true space-time with a fluctuating geometry. Another question is whether there is a deeper symmetry behind this Poincar\'e invariance, with for instance field transformations coupling all arguments of the group field and not simply acting on pairs of arguments. We could also investigate if the Poincar\'e invariance is somehow related  to the translational symmetry of topological BF theory (see e.g. \cite{PR1}).

Another interesting point is to check the symmetries for the new type of group field theories, introduced in \cite{newgft}, where the field have both $g$ and $B$ arguments. Possibly the deformed Poincar\'e symmetry would have a clearer geometrical interpretation in that framework.

Finally, we should further look into the four-dimensional case and investigate the classical symmetries  of group field theories corresponding to non-topological theories. This would help to understand which symmetry we should require for a group field theory properly quantizing gravity.

\section*{Acknowledgments}

EL is partially supported by the ANR ``Programme Blanc" grants LQG-09.
%



\end{document}